# Instantaneous identification of Bouc-Wen-type hysteretic systems from seismic response data

R. Ceravolo[1,a], G. V. Demarie[1,b] and S. Erlicher[2,c]

[1] Dip. Ingegneria Strutturale e Geotecnica, Politecnico di Torino, Turin Italy

[2] LAMI-ENPC, Champs sur Marne, France

[a]rosario.ceravolo@polito.it, [b]giacomo.demarie@polito.it, [c]erlicher@lami.enpc.fr

**Keywords:** hysteretic models, structural identification, instantaneous estimators, seismic response

**Abstract.** This paper presents a technique for identification of non-linear hysteretic systems subjected to non-stationary loading. In the numerical simulations, a Bouc-Wen model was chosen for its ability to represent the properties of a wide class of real hysteretic systems. The parameters of the model are computed instantaneously by approximating the internal restoring force surface through an "ad hoc" polynomial basis. Instantaneous estimates result from time-varying spectra of the response signals. A numerical application of interest to earthquake engineering is finally reported.

## Introduction

Methods for identifying systems with hysteresis can be framed respectively in the *parametric* and the *non parametric* approach: in the former case, *a priori* selection of a specific model for the dynamic behaviour of the system is needed and the identification process consists of determining the coefficients for such model. Non parametric methods, instead, do not require any assumption on the type and localisation of structural non-linearities but, generally, the quantities identified cannot be directly correlated to the system's equation of motion and, therefore, their use in the applications is limited.

Chassiakos et al. [1] and Smyth et al. [2] proposed a parametric method in which the parameters of the Bouc-Wen model are identified through an adaptive procedure, based on the application of least square techniques of estimation. An alternative approach is due to Kyprianou et al [3], who introduced a "differential evolutive" method for the identification of the parameters of the Bouc-Wen model, whose formulation in many respects comes close to that of genetic algorithms.

The formulation of parametric methods requires the adoption of an appropriate form for the model, which should accurately describe the dynamic behaviour of the system under investigation. Sometimes, however, it may prove impossible to meet this requirement: this is what prompted the development of non-parametric methods whose formulation does not require any a priori knowledge of the dynamics of a structure. Classical non-parametric methods are based on the extension of the restoring force surface method: Benedettini et al. [4] approximated the surface of the time derivative of the internal restoring force on a polynomial basis, by assuming as state variables the force itself and velocity; Masri et al. [5] extended this approach by proposing a polynomial base approximation of the system as a function of velocity, displacement and the excitation.

In the frame of non-parametric approaches, Pei et al. [6] used a special type of neural network, which showed good performances in the identification of hysteretic systems. Finally, Saadat et al. [7] formulated a "hybrid" approach that combine in a single identification procedure the potentials of both the parametric and the non parametric approach.

The present paper adopts the idea of approximating the restoring force surface on a polynomial basis [4,5] and outlines a strategy for an instantaneous identification of the hysteretic model parameters.

## Bouc-Wen hysteretic models

The Bouc model formulation is based on the use of the following Stieltjes integral [8]:

$$r(t)=\int_0^{\xi(t)} \mu\left(\xi(t)-\xi'\right)dx(\xi') \qquad (1)$$

where $x$ and $r$ are two time-dependent functions, which are considered as input and output functions, respectively. In structural engineering applications the input $x$ usually has the meaning of a generalized displacement, while the output $r$ plays the role of a generalized force. The integral in Eq. 1 depends on the time-function $\xi(t)$, which is referred to as internal time and is assumed to be positive and non-decreasing. The function $\mu$, called the hereditary kernel, takes into account hysteretic phenomena. One of the definitions of $\xi(t)$ proposed by Bouc, is the total variation of $x$:

$$\xi(t)=\int_0^t \left|\frac{dx}{dt'}\right| dt' \quad \Leftrightarrow \quad \dot{\xi}=|\dot{x}|, \quad with\ \xi(0)=0 \qquad (2)$$

where the superposed dot indicates time differentiation. Eq. 2 implies the rate-independence of $\xi(t)$ and as a result, $r(t)$ is in turn rate-independent. The kernel $\mu$ is defined as a continuous, bounded, positive and non-increasing function on the interval $\xi(t) \geq 0$, having a bounded integral. In the special case of an exponential kernel $\mu(\xi)=Ae^{-\beta\xi}$, with $A,\beta>0$, a differential formulation of Eq. 1 can be easily deduced:

$$\dot{r}=A\dot{x}-\beta r\dot{\xi} \qquad (3)$$

with $\dot{\xi}=|\dot{x}|$. One can observe that for an initial value in the interval ($-r_u$, $r_u$), with $r_u = A/\beta$, the hysteretic force $r(t)$ remains in the same interval. The univariate (1D) Bouc model of Eq. (3) was modified by the contributions of several authors; among others, Wen suggested the use of the positive exponent, $n$:

$$\dot{r}=A\dot{x}-\left(\beta sign|r\,\dot{x}|+\gamma\right)|r|^n\,\dot{x} \qquad (4)$$

where sign(. ) is the signum function. Note that Eq. 4 can be written under the form (3), with the intrinsic time flow $\dot{\xi}=\left(1+sign(r\,\dot{x})\gamma/\beta\right)|r|^{n-1}|\dot{x}|$. Wen assumed integer values for $n$; however, all real positive values of $n$ are admissible. When $n$ is large enough, force-displacement curves similar to those of an elastic-perfectly-plastic model are obtained. Provided that $\beta+\gamma > 0$, the limit strength value $r_u$ of the model of Eq. 4 becomes $r_u=\sqrt[n]{A/(\beta+\gamma)}$. The parameter $\beta$ is positive by assumption, while the admissible values for γ, i.e. $\gamma\in[-\beta,\beta]$, can be derived from a thermodynamic analysis [9].

A further important modification concerned the introduction of the so-called strength and stiffness degradation effects, by means of the functions $\eta$ and $\nu$ [10]:

$$\dot{r}=\frac{1}{\eta}\left[A\dot{x}-\nu\, r\,\beta\,\dot{\xi}\right] \qquad (5)$$

where $\dot{\xi}$ is the same as for a non-degrading model; $\nu = 1 + c_\nu e$ represents a strength degradation effect, while $\eta = 1 + c_\eta e$ is associated with a stiffness degradation effect; *e* is the energy dissipated by the hysteretic model and $c_\nu \geq 0$ and $c_\eta \geq 0$.

The identification technique will be explained referring to the case of a simple Bouc-Wen oscillator, whose equation of motion is:

$$\begin{cases} m\ddot{x} + r = u \\ \dot{r} = A\dot{x} - \left[\beta \operatorname{sgn}(r \cdot \dot{x}) + \gamma\right] \cdot |r|^n \dot{x} \end{cases} \tag{6}$$

where *m*, $\ddot{x}$ and *u* are respectively the mass, the acceleration and the excitation, *r* is the internal restoring force; *A*, $\beta$, $\gamma$ and *n* are the parameters on which the form of the hysteresis cycle depends.

For the purposes of identification, the function $\dot{r}(\dot{x}, r)$ is replaced with a polynomial in the variables $\dot{x}$ and *r*:

$$\begin{bmatrix} m & 0 \\ 0 & 0 \end{bmatrix} \begin{Bmatrix} \ddot{x} \\ \ddot{r} \end{Bmatrix} + \begin{bmatrix} 0 & 0 \\ -\alpha_1 & 1 \end{bmatrix} \begin{Bmatrix} \dot{x} \\ \dot{r} \end{Bmatrix} + \begin{bmatrix} 0 & 1 \\ 0 & -\alpha_2 \end{bmatrix} \begin{Bmatrix} x \\ r \end{Bmatrix} + \begin{Bmatrix} 0 \\ -\alpha_3 \dot{x}^3 - \alpha_4 \dot{x}^2 r - \alpha_5 \dot{x} r^2 - \alpha_6 r^3 - \ldots \end{Bmatrix} = \begin{Bmatrix} u \\ 0 \end{Bmatrix} \tag{7}$$

where $\alpha_1$, $\alpha_2 \ldots$ are the coefficients of the polynomial approximation of $\dot{r}$. Within the range of validity of the approximation adopted, the above equation admits Volterra series expansion.

**Instantaneous estimators of Bouc-Wen parameters**

Let $\{p\}^T = \mathbf{p}^\mathbf{T} = \{m \quad \alpha_1 \quad \alpha_2 \quad \alpha_3 \quad \alpha_4 \quad \ldots\}$ be the vector of parameters which describes the dynamic properties of the system and let us introduce an instantaneous objective function [11,12]:

$$F_{ob}\left(n^*, \mathbf{p}\right) = \sum_{m=0}^{M-1} \left| \left| D\left(n^*, m\right) \right|^2 - \left| D_V\left(n^*, m, \mathbf{p}\right) \right|^2 \right| \tag{8}$$

where $D\left(n^*, m\right)$ and $D_V\left(n^*, m, \mathbf{p}\right)$ are the values of the response signal's time-dependent spectrum, e.g. a Short-Time Fourier Transform (STFT) [13], at discrete time instant $n^* \Delta t$ and frequency $m \Delta f$, respectively measured and calculated by integrating Eq. 7, for a given configuration of parameters $\mathbf{p}$. The number of frequency samples, *M*, to be considered in the objective function in Eq. 8 will depend on the frequency interval under analysis.

If the time-dependent spectrum satisfies the time marginal property [13], then $F_{ob}$ gives the difference between the instantaneous energies of the experimental signals and those of the system output corresponding to a given configuration of the unknown parameters $\mathbf{p}$: by resorting to classical optimisation procedures, it proves possible to determine instant by instant the minimum of $F_{ob}\left(n^*, \mathbf{p}\right)$, through which the vector of the instantaneous estimators $\mathbf{p}(t)$ of the dynamic properties can be defined.

The knowledge of the instantaneous estimators $\alpha_1(t)$, … $\alpha_4(t)$, as determined at time $t = n^* \cdot \Delta t$, makes it possible to obtain an approximation of the surface $\dot{r}$ which is valid in a portion of the definition domain comprised between the maximum values that the velocity and the internal restoring force of the system assume around time *t*.

The analytical expression of surface $\dot{r}$ (Eq. 6) depends linearly on the *A*, $\beta$ and $\gamma$ constants, and non-linearly on the *n* exponent; if a value of *n* is selected a priori, then the instantaneous estimators $A(t)$, $\beta(t)$ and $\gamma(t)$ for the Bouc-Wen model can be determined by solving at each time a linear system with the classical least squares techniques. The limits of the region of the $(\dot{x}, r)$ where linear equations are written can be defined on the basis of the values taken on by the instantaneous amplitude of the velocity, $\dot{x}$, and of the internal restoring force, *r*, assumed around the instant at which the instantaneous estimators are evaluated.

## Numerical example

As a numerical application of the instantaneous identification, a Single Degree Of Freedom (SDOF) system described by the Bouc-Wen model subjected to a seismic excitation at the support (San Fernando earthquake, 1971, E-W component, accelerations measured on the basement of the Alhambra Building, sampling frequency: 200 Hz [14]) has been considered.

The structural response used for the identification of system parameters was the relative acceleration, as estimated through step integration of Eq. 7 using the ©SIMULINK software. Figs. 1 and 2 represent the system response and a diagram of the internal restoring force as a function of displacement, Table 1 gives the numerical values of the Bouc-Wen model parameters.

The identification was performed in the 1-80 s time interval and by considering the presence, in the "measured response" of exogenous noise produced by contaminating the acceleration with white Gaussian noise having an RMS corresponding to 10% the RMS value of system acceleration; moreover, the value of mass *m* was supposed to be known a priori.

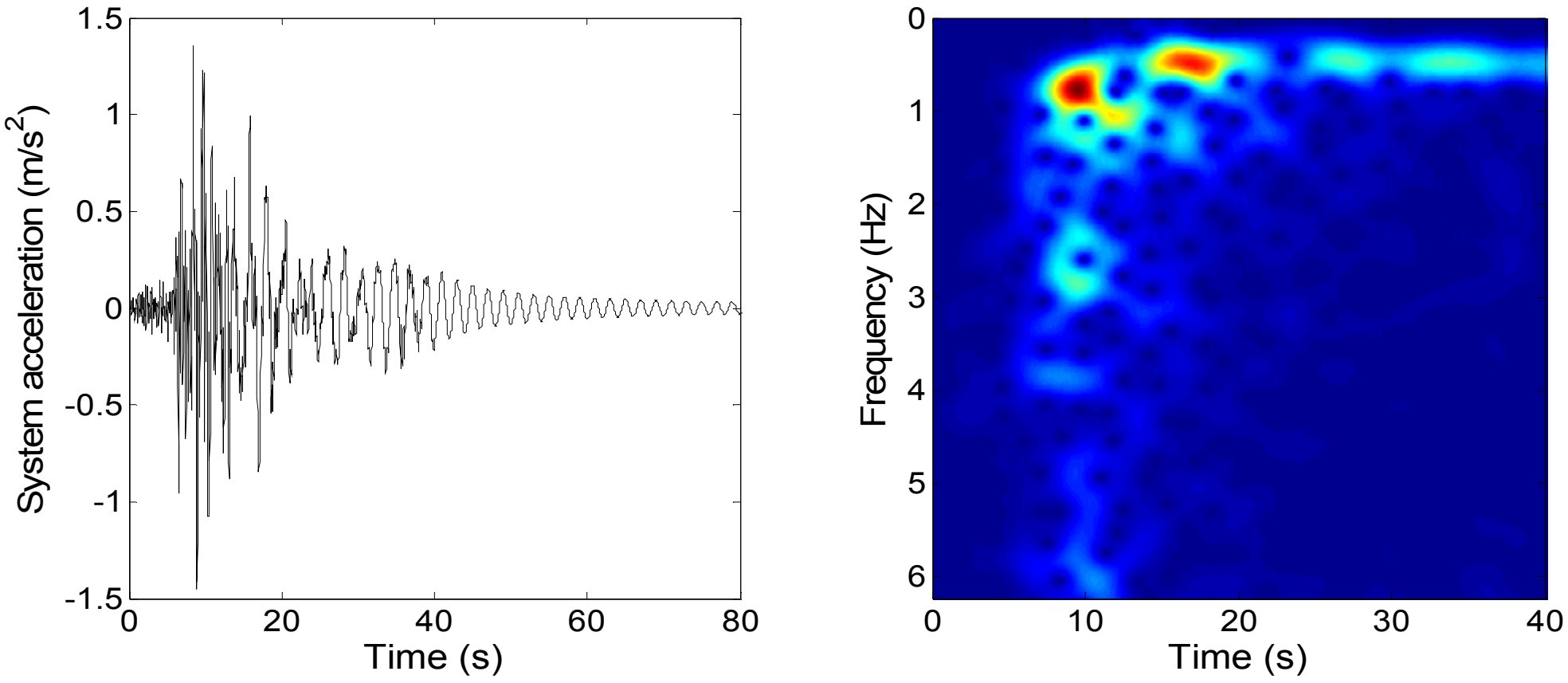

Figure 1: Bouc-Wen oscillator: system output and a Gabor *t-f* representation

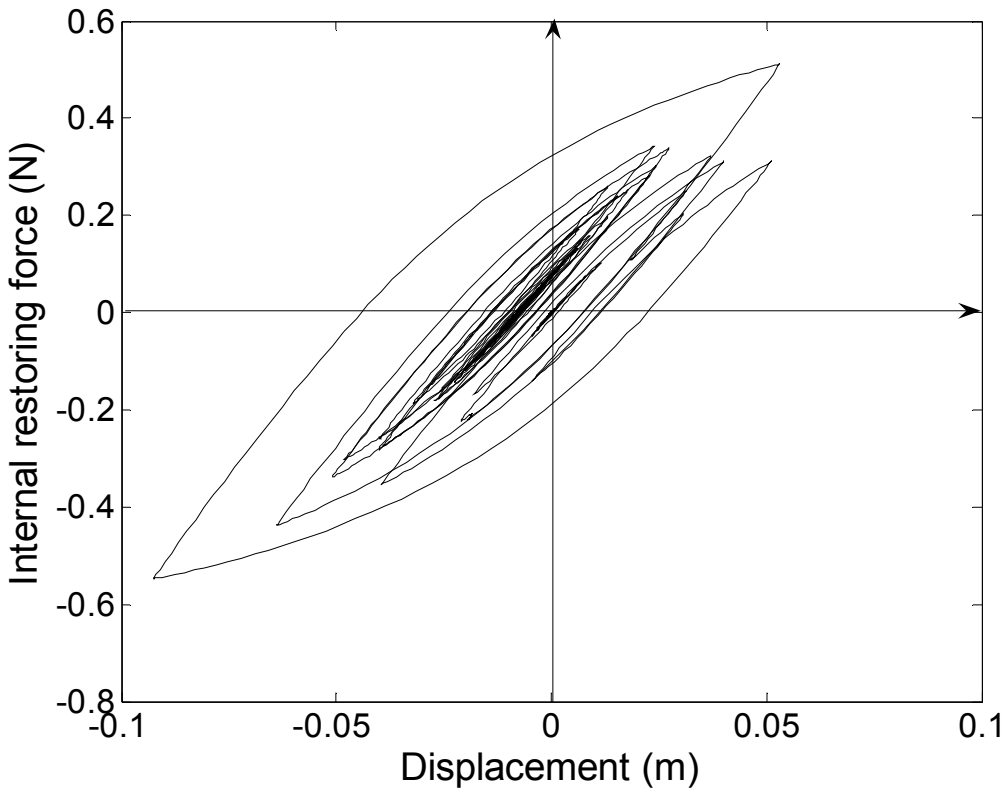

Figure 2: Bouc-Wen oscillator: system output - hysteretic loop

Table 1: System parameters

| | |
|---|---|
| $m$ | 1 kg |
| $A$ | 10 N/m |
| $\beta$ | 10 $N^{(1-n)} \cdot m^{-1}$ |
| $\gamma$ | 5 $N^{(1-n)} \cdot m^{-1}$ |
| $n$ | 1 |

Eq. 7 can be expanded in Volterra series, by resorting to associated linear equations (ALEs) [15], the first of which is the equation of motion of the underlying linear system; having posed the Wen parameter $n$=1, the best cubic polynomial was found to contain only the $\dot{x}^2 r$ and $\dot{x} r^2$ terms, hence:

$$\begin{cases} [M]\{\ddot{q}\}_1 + [C]\{\dot{q}\}_1 + [K]\{q\}_1 = \{u\} \\ [M]\{\ddot{q}\}_3 + [C]\{\dot{q}\}_3 + [K]\{x\}_3 = \begin{Bmatrix} 0 \\ \alpha_4 \dot{x}_{(1)}^2 r_{(1)} + \alpha_5 \dot{x}_{(1)} r_{(1)}^2 \end{Bmatrix} \\ [M]\{\ddot{q}\}_5 + [C]\{\dot{q}\}_5 + [K]\{x\}_5 = \begin{Bmatrix} 0 \\ \alpha_4 \left(\dot{x}_{(1)}^2 r_{(3)} + 2\dot{x}_{(1)} \dot{x}_{(3)} r_{(1)}\right) + \alpha_5 \left(2\dot{x}_{(1)} r_{(1)} r_{(3)} + \dot{x}_{(3)} r_{(1)}^2\right) \end{Bmatrix} \\ [M]\{\ddot{q}\}_7 + [C]\{\dot{q}\}_7 + [K]\{x\}_7 = \begin{Bmatrix} 0 \\ \alpha_4 \left(\dot{x}_{(1)}^2 r_{(5)} + 2\dot{x}_{(1)} \dot{x}_{(5)} r_{(1)} + 2\dot{x}_{(1)} \dot{x}_{(3)} r_{(3)} + \dot{x}_{(3)}^2 r_{(1)}\right) + \\ + \alpha_5 \left(\dot{x}_{(5)} r_{(1)}^2 + 2\dot{x}_{(1)} r_{(1)} r_{(5)} + 2\dot{x}_{(3)} r_{(1)} r_{(3)} + \dot{x}_{(1)} r_{(3)}^2\right) \end{Bmatrix} \\ \ldots \end{cases} \tag{9}$$

$$\{q\} = \{q\}_1 + \{q\}_3 + \{q\}_5 + \{q\}_7 + \ldots$$

being $\dot{x}_{(s)}$ $r_{(s)}$ the velocity and the internal force corresponding to the $n$-th component of the Volterra series expansion of the system's response; [$M$], [$C$] and [$K$] are the matrices of the linear part of the system. From Eq. 9 it is possible to determine the $s$-th order contribution of the Volterra series expansion, $\{q(t)\}_s$, as the response of the underlying linear system to an input which is a non-linear function of the first $s$-1 terms.

A proper choice for the order of the Volterra series expansion has the advantage of suppressing distortions due to measurement or integration; in this case the order 7 was found to be sufficient to adequately approximate the system response.

Fig. 3 illustrates the instantaneous estimators of the coefficients of the equivalent polynomial system: the evolution of the non-linear parameters $\alpha_4$ and $\alpha_5$ shows a high variability in the interval 10-40 s, where the system response is clearly non-linear. Outside this interval the instantaneous estimators do not present any fluctuations, the output being essentially linear.

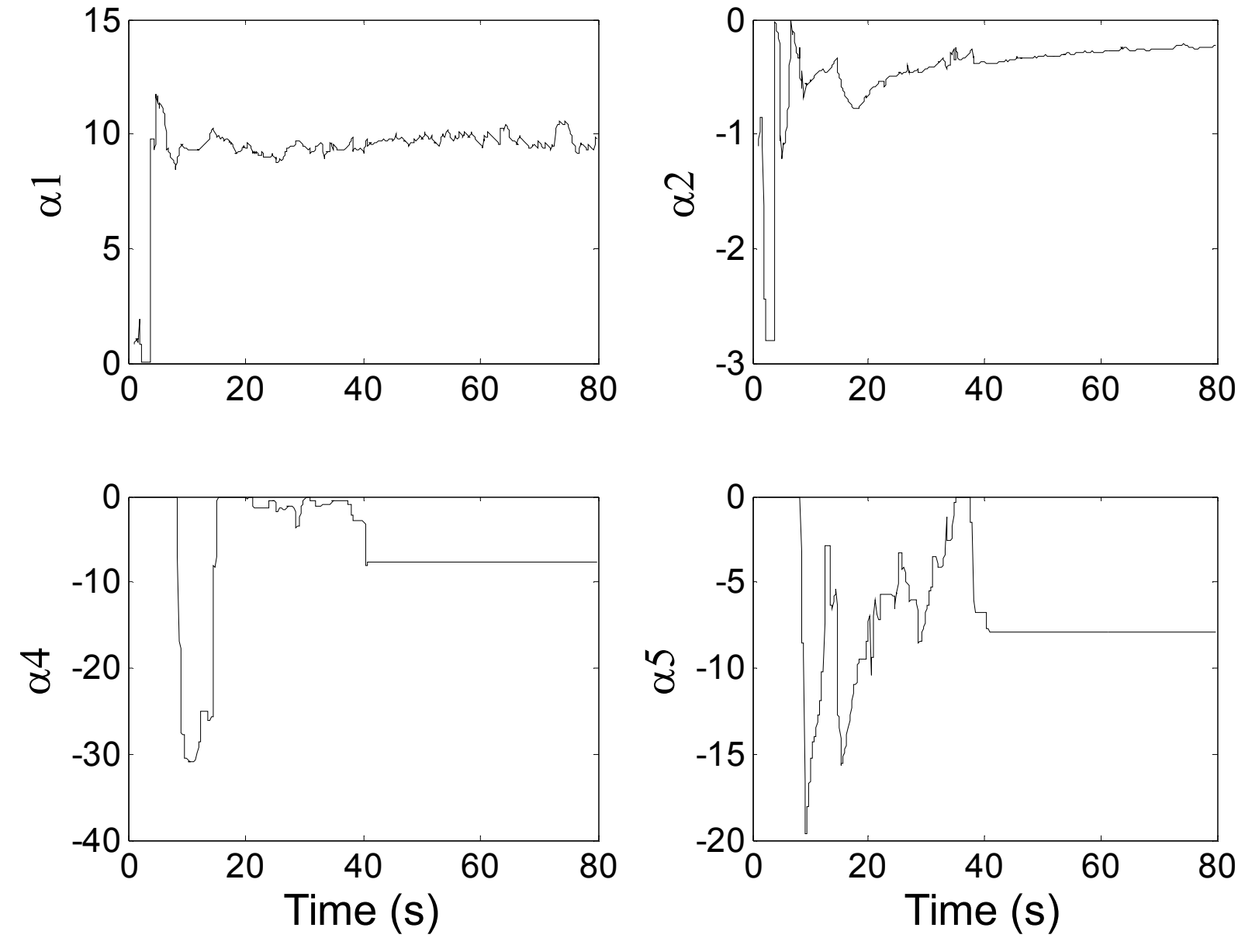

Figure 3: Instantaneous estimators of the equivalent polynomial system

The identified acceleration was evaluated by subdividing the total time support into smaller intervals (2s) and calculating, over the small interval considered, the response of the equivalent polynomial system whose coefficients are the mean of the instantaneous estimators. The excellent agreement between the time histories (Fig. 4a) evidences the positive result of the identification procedure and the ability of the equivalent polynomial model to reproduce correctly the dynamic behaviour of the system in the portion of domain identified by the maximum values of $|\dot{x}|$ and $|r|$ observed during the vibration. The quality of the instantaneous approximation is related to the length of the analysis window in samples, whose optimal value in its turn depends on decorrelation length [11,13]: the diagram in Fig. 4b shows the error in the equivalent polynomial approximation as a function of the STFT window length in samples for this specific application.

Fig. 5 shows the instantaneous estimators of the Bouc-Wen model coefficients, evaluated in the maximum amplitude range for velocity and the internal restoring force (10-20 s). Table 2 lists the mean characteristics of the coefficients identified (STFT window length: 200 samples).

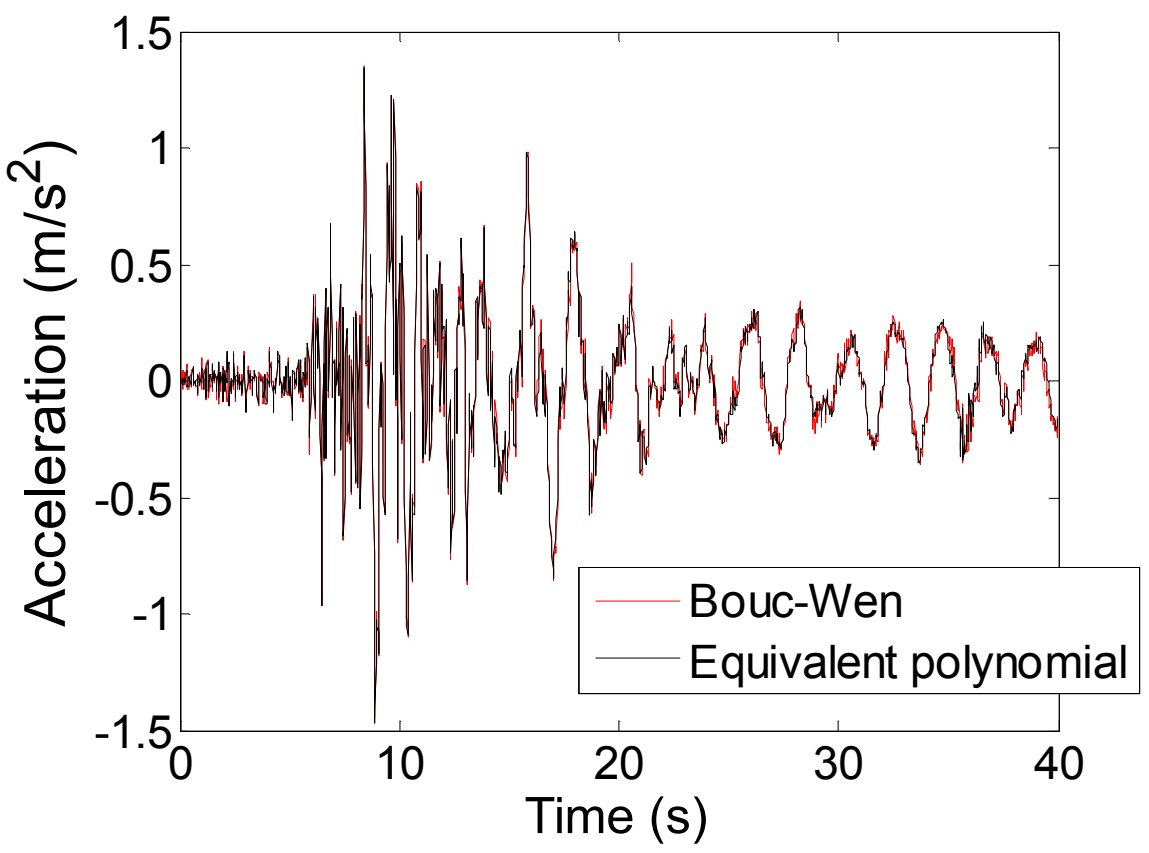

Figure 4a: Comparison between the "measured" response, "Bouc-Wen", and the identified one, "equivalent polynomial" (window length: 201 samples)

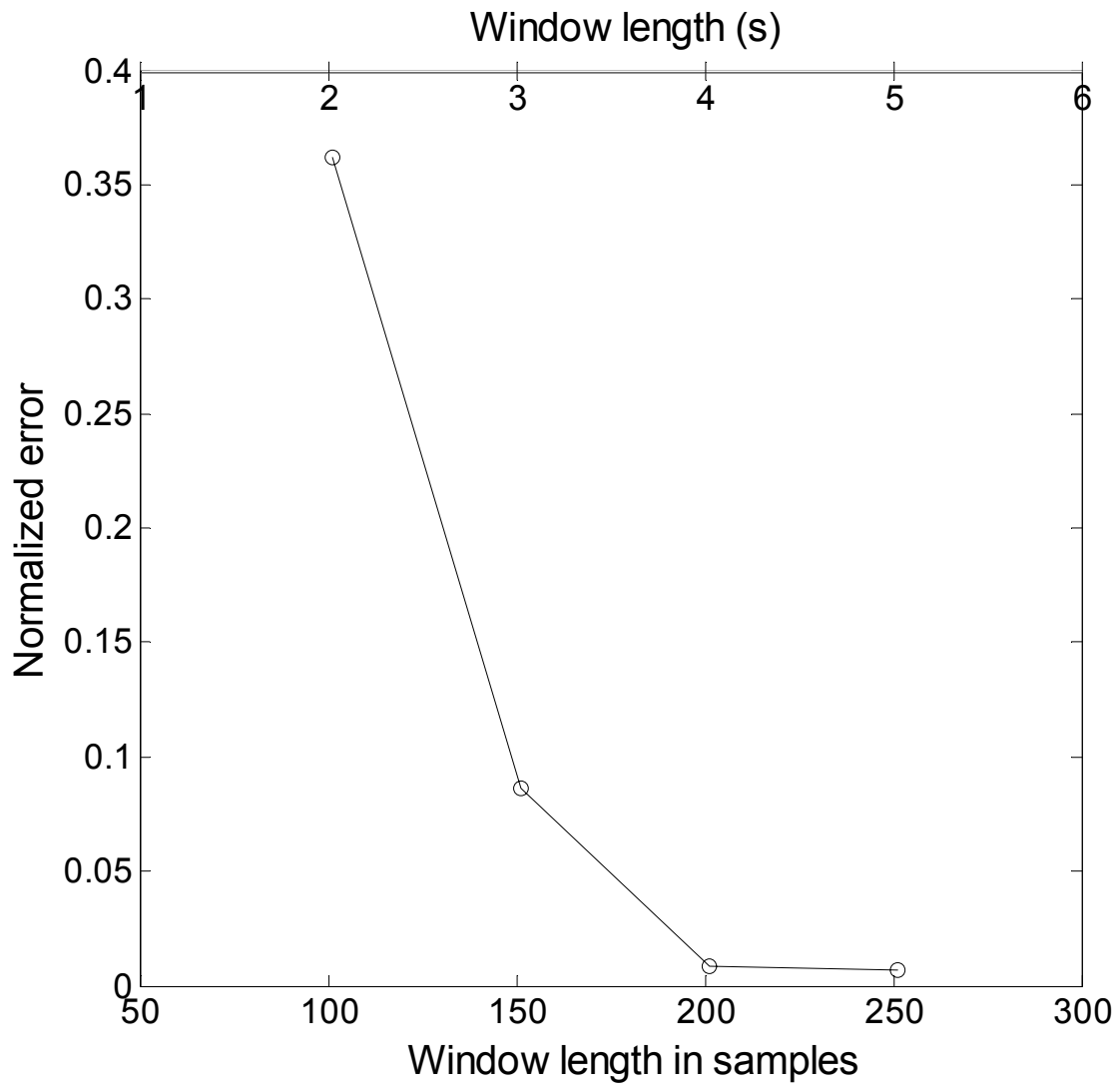

Figure 4b: Error in the equivalent polynomial approximation as a function of window length in samples (as normalised with respect to the signal's energy)

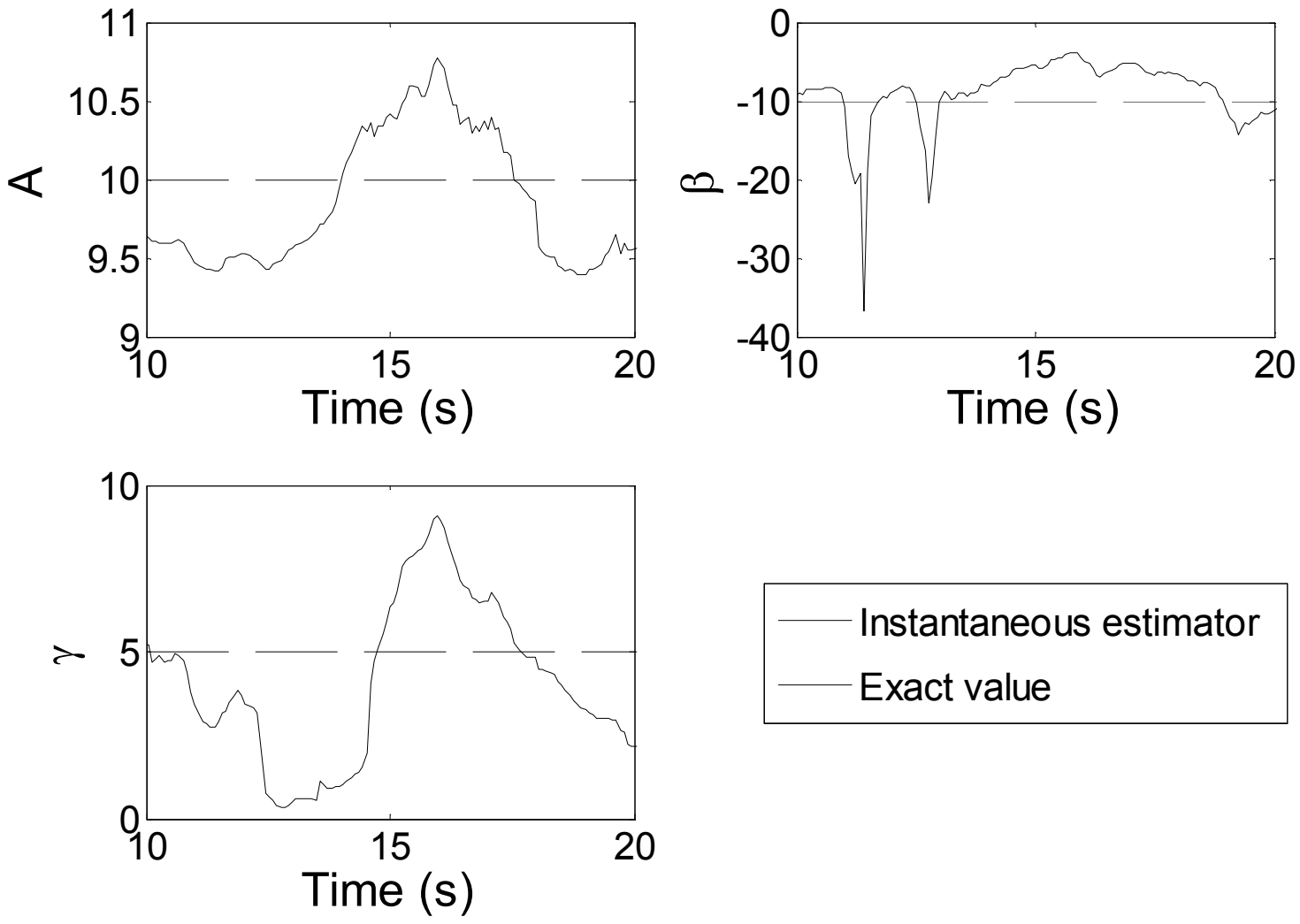

Figure 5: Instantaneous estimators for the Bouc-Wen model

Table 2: Mean characteristics of the instantaneous estimators

| Parameter | Exact value | Mean | Standard deviation | Coefficient of variation |
|---|---|---|---|---|
| $A$ | 10 | 9.84 | 0.40 | 0.041 |
| $\beta$ | 10 | 8.96 | 4.09 | 0.457 |
| $\gamma$ | 5 | 4.21 | 2.34 | 0.555 |

The instantaneous estimator of coefficient $A$ is characterised by considerable stability around the exact value, whilst the scatter in the coefficients of the non-linear part of the Bouc-Wen model is significantly higher, as borne out by the significant magnitude of the standard deviations and the coefficients of variation. The identification of these parameters is affected not solely by the presence of external noise; in the time intervals where the excitation and the dynamic response are modest, in fact, system behaviour can be rated as linear and the determination of the $\beta$ and $\gamma$ coefficient is not reliable. Fig. 6 shows in the $(\dot{x}, r)$ plane the internal restoring force surfaces obtained by substituting into the analytical expression (Eq. 6) the exact values and the mean values of the instantaneous estimates (Tab. 1 and Tab. 2, respectively). This figure confirms that the approximation obtained, relative to the portion of domain identified by the maximum values of $|\dot{x}|$ and $|r|$, is acceptable and, hence, the result of the identification can be regarded as positive.

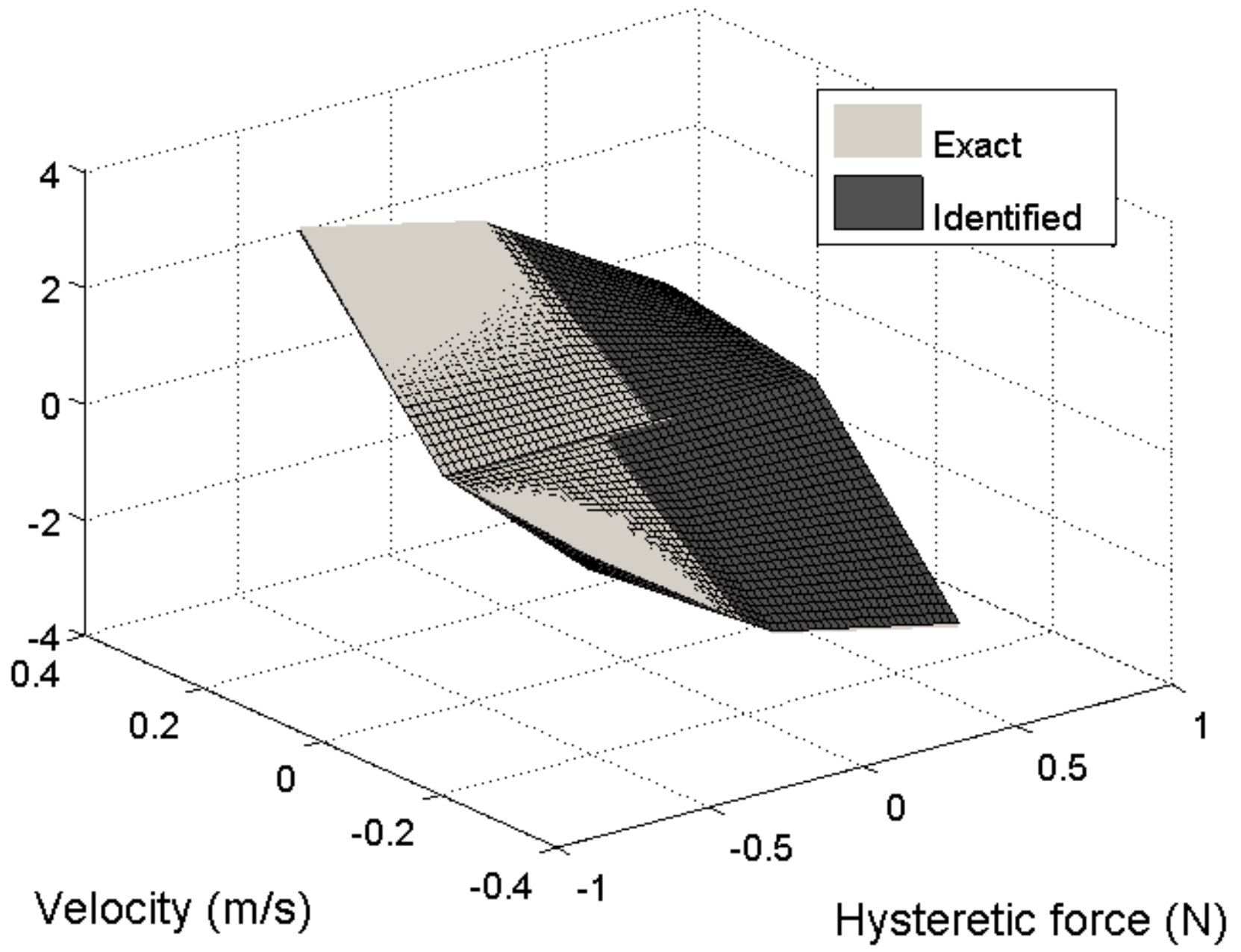

Figure 6: Restoring force surface: comparison between the exact surface and the surface identified

## Conclusions

This paper has presented a parametric procedure for the identification of the parameters of hysteretic systems described by the Bouc-Wen model. The determination of the model parameters has been conducted by approximating the surface of the time derivative of the internal restoring force, $\dot{r}$, with a cubic polynomial in the $\dot{x}$ and $r$ variables and calculating the dynamic response of the system through a finite Volterra series.

The application to a Bouc-Wen SDOF subjected to a seismic excitation has lead to the identification of an equivalent time-variant polynomial system, from which it was possible to get instantaneous estimates of the system parameters.

In addition to reducing computational costs, the use of Volterra series expansion demonstrated to be advantageous in suppressing distortions due to measurement or integration, which otherwise would have been attributed to higher-order Volterra kernels; in the numerical example, the order 7 was found to be sufficient to adequately approximate the system response.